\documentclass{LMCS}
\usepackage{latexsym,amsmath,amssymb,hyperref}

\renewcommand{\phi}{\varphi}
\newcommand{\twentieth}{\ensuremath{20^\mathrm{th}}}

\def\doi{2 (3:6) 2006}
\lmcsheading%
{\doi}
{1--13}
{}
{}
{Feb.~24, 2006}
{Sep.~27, 2006}
{}

\begin{document}

\bibliographystyle{plain}

\title{Tarski's influence on computer science}
\author{Solomon Feferman}
\address{Departments of Mathematics and Philosophy\\
         Stanford University}
\email{sf@csli.stanford.edu}
\titlecomment{}  
 
\begin{abstract}
\noindent
Alfred Tarski's influence on computer science was indirect but significant in a number of directions and was in certain respects fundamental.  Here surveyed is Tarski's work on the decision procedure for algebra and geometry, the method of elimination of quantifiers, the semantics of formal languages, model-theoretic preservation theorems, and algebraic logic; various connections of each with computer science are taken up.
\end{abstract}

\maketitle

The following is the text of an invited lecture for the LICS 2005
meeting held in Chicago June 26-29, 2005.%
\footnote{I want to thank the
organizers of LICS 2005 for inviting me to give this lecture and for
suggesting the topic of Tarski's influence on computer science, a
timely suggestion for several reasons. I appreciate the assistance of Deian Tabakov and Shawn Standefer in preparing the \LaTeX version of this text. 
Except for the addition of
references, footnotes, corrections of a few points and stylistic
changes, the text is essentially as delivered.  Subsequent to the
lecture I received interesting comments from several colleagues that
would have led me to expand on some of the topics as well as the list
of references, had I had the time to do so. }

Almost exactly eight years ago today, Anita Feferman gave a lecture
for LICS 1997 at the University of Warsaw with the title, ``The saga
of Alfred Tarski: From Warsaw to Berkeley.'' Anita used the
opportunity to tell various things we had learned about Tarski while
working on our biography of him.  We had no idea then how long it
would take to finish that work; it was finally completed in 2004 and
appeared in the fall of that year under the title, \emph{Alfred Tarski: Life
and Logic} \cite{feferman04a}.  The saga that Anita recounted took Tarski from the
beginning of the \twentieth{} century with his birth to a middle-class
Jewish family and upbringing in Warsaw, through his university studies
and Ph.D. at the ripe young age of 23 and on to his rise as the
premier logician in Poland in the 1930s and increasing visibility on
the international scene--despite which he never succeeded in obtaining
a chair as professor to match his achievements.  The saga continued
with Tarski coming to Harvard for a meeting in early September, 1939
when the Nazis invaded Poland on September 1st, at which point he was,
in effect, stranded.  Then, during the next few years he went from one
temporary research or teaching position to another on the East Coast.
He was finally offered a one year position in 1942 as Lecturer in
Mathematics at the then far off University of California in Berkeley,
with the suggestion that it might stretch into something longer.  In
fact, he not only succeeded in staying, but rose to the rank of
Associate Professor by the end of the war and a year later was made
Full Professor, thus finally obtaining the position he deserved.  At
Berkeley, Tarski built from scratch one of the world's leading
centers in mathematical logic, and he remained there, working
intensively with students, colleagues and visitors until his death in
1983.

Tarski became recognized as one of the most important logicians of the
\twentieth{} century through his many contributions to the areas of
set theory, model theory, the semantics of formal languages, decidable
theories and decision procedures, undecidable theories, universal
algebra, axiomatics of geometry, and algebraic logic.  What, in all
that, are the connections with computer science?  When Anita started
working on the biography--which only later became a joint project--she
asked me and some of my colleagues exactly that question, and my
response was: none.  In contrast to that--as she said at the
conclusion of her Warsaw lecture--John Etchemendy (my colleague in
Philosophy at Stanford, and now the Provost of the University)
responded: ``You see those big shiny Oracle towers on Highway 101?
They would never have been built without Tarski's work on the
recursive definitions of satisfaction and truth.''%
\footnote{For those who may not know what the ``big shiny Oracle towers'' are, the reference is to the headquarters of Oracle Corporation on the Redwood Shores area of the San Francisco Peninsula.   A duly shiny photograph of a few of these towers may be found at \goodbreak
\texttt{http://en.wikipedia.org/wiki/Image$:$Oracle$\_$Corporation$\_$HQ.png}.} It took me a while
to see in what sense that was right.  Indeed--as I was to learn--there
is much, much more to say about his influence on computer science, and
that's the subject of my talk today.  I owe a lot to a number of
colleagues in the logic and computer science areas for pointing me in
the right directions in which to pursue this and also for providing me
with very helpful specific information.\footnote{I am most indebted in
this respect to Phokion Kolaitis.  Besides him I have also received
useful comments from Michael Beeson, Bruno Buchberger, George Collins,
John Etchemendy, Donald Knuth, Janos Makowsky, Victor Marek, Ursula
Martin, John Mitchell, Vaughan Pratt, Natarajan Shankar, and Adam
Strzebonski.  And finally, I would like to thank the two anonymous referees for a number of helpful corrections.}
	
\emph{Alfred Tarski: Life and Logic} was written for a general
audience; the biographical material is interspersed with interludes
that try to give a substantive yet accessible idea of Tarski's main
accomplishments.  Still, given the kind of book it is, we could not go
into great detail about his achievements, and in particular could only
touch on the relationship of his work to computer science.  Before
enlarging on that subject now, I want to tell a story that \emph{is}
in our biography (\cite{feferman04a}, p. 220--230),
and is in many respects revelatory of his own attitude
towards the connection.

I had the good fortune to be Tarski's student in the 1950s when he
was beginning the systematic development of model theory and algebras
of logic.  In 1957, the year that I finished my Ph.D., a month-long
Summer Institute in Symbolic Logic was held at Cornell University.
That proved to be a legendary meeting; in the words of Anil Nerode: 
``There has been nothing else in logic remotely comparable.'' What the
Cornell conference did was to bring together for the first time,
leaders, up-and-coming researchers, and students in all the main areas
of logic, namely model theory, set theory, recursion theory, and proof
theory.  Besides Tarski, the top people there--along with their
coteries--were Alonzo Church, Stephen Kleene, Willard Quine, Barkley
Rosser, and, in the next generation, Abraham Robinson and George
Kreisel.  The organization of the meeting itself had been inspired by
the mathematician Paul Halmos, who, independently of Tarski, had
developed another approach to the algebra of first-order logic.  As
Halmos wrote about it in his  {\em Automathography} \cite{halmos85}, p. 215:
\begin{quote}
\small    
There weren't many conferences, jamborees, colloquia in those days
and the few that existed were treasured. .... I decided it would be
nice to have one in logic, particularly if it were at least partly
algebraic.
\end{quote}
And, ``nice'' it was.  

In addition to the four main areas, the Cornell logic conference was
the first to include many speakers from the emerging field of computer
science, the theoretical foundations of which had been laid in the 30s
by G\"odel, Church, Turing, Post, and Kleene.  The connections between
the theory and application of computation began toward the end of
World War II when the first large scale electronic digital computers
were built.  At that point, for each kind of application, the hardware
had to be programmed by hand, a long and arduous task. John von
Neumann was instrumental in demonstrating how to circumvent that process by introducing the
first form of software.  

By 1957, companies such as IBM and Remington Rand were producing the
first generation of commercial electronic computers, and the
high-level programming language FORTRAN had become established as an
industry standard.  Some--but by no means all--logicians were quick to
grasp the implications of these developments.  At the Cornell meeting,
Rosser gave a talk on the relation between Turing machines and actual
computers; Church gave a series of talks on the logical synthesis of
switching circuits for computer hardware; and Abraham Robinson spoke
on theorem proving as done by man and machine.  Among the younger
contributors, Michael Rabin and Dana Scott spoke about finite
automata, and Martin Davis talked about his implementation on the
``Johnniac'' computer (at the Institute for Advanced Study) of a
decision procedure for the arithmetic of the integers under
addition--a procedure that had been discovered in 1930 by Tarski's
student Mojzesz Presburger in his Warsaw seminar.  

On the industry side, IBM and some of the other companies employed a
number of researchers with backgrounds in mathematics and logic, and
these people turned out in large numbers at Cornell, both to listen
and to speak.  There were fifteen talks given by researchers from IBM,
many of them demonstrating the utility of FORTRAN-like programs for
solving problems of potential interest to logicians.  In particular,
the talk by George Collins--a former student of Rosser's--on the
implementation of parts of Tarski's decision procedure for the algebra
of real numbers on an IBM 704 should have caught Tarski's attention
because it suggested possible practical applications of his procedure.
But a few years ago, when I asked Collins about Tarski's reaction to
that, he said: ``He didn't show any appreciation for my work, either
then or later.  I was somewhat surprised and disappointed.''  It is
indeed surprising that--despite Tarski's own recognition of the
importance and systematic pursuit of the decision problem for various
algebraic theories, he did not evince the least bit of interest in the
practical computational applications of those problems for which a
decision procedure had been found.  And, he didn't even seem to be
interested either in the work of Rabin and Scott, both of whom were
high on his list of favorites, Scott as a former student and
collaborator, and Rabin as someone he wanted to bring to Berkeley.

Let's look at what Tarski and Collins were up to in more detail.  At
heart, Tarski's decision procedure rests on the solution of an
algebraic problem for the reals, i.e., the ordered field $\langle
\mathbb{R},+,\times,<, 0,1 \rangle$.  Tarski's procedure uses the method of
elimination of quantifiers to associate with each first-order formula
of the language of the reals an equivalent quantifier-free formula;
those without free variables are then easily decided.  The procedure
reduces to determining for each system $P$ of polynomial equations and
inequalities and one of its variables $x$, whether or not there exists
a common real solution $x$; the answer is to be expressed in terms of
the coefficients of the polynomials involved and the remaining
variables.  Algorithms for special cases of this problem go back
through the history of algebra.  Tarski's procedure generalizes one
due to Sturm for computing the number of roots of a real polynomial in
a given interval.  On the face of it, Tarski's procedure is
non-elementary in time complexity, i.e., greater than all finite towers
of powers of 2, and so it was important for Collins to find a more
feasible procedure than the one that he had talked about at the 1957
Cornell meeting.  Further improvements on Tarski's procedure by
Abraham Seidenberg and later by Paul Cohen didn't really help much in
that respect.  In the meantime, Collins was working on various aspects
of computer algebra and in 1974 and 1975 he published \cite{collins74}, \cite{collins75} a new method of
doubly exponential upper bound complexity called Cylindrical Algebraic
Decomposition (CAD).  Incidentally, this was during a year that he was
visiting Stanford University from Wisconsin.\footnote{Collins reports
(\cite{caviness98}, p. 86) a communication from Leonard
Monk in 1974 stating that he and Bob Solovay had obtained a triply
exponential upper bound decision procedure for real algebra, though
not a quantifier elimination procedure.  Fischer and Rabin say
(\emph {op. cit.}, p. 124) that Solovay found a doubly exponential upper
bound, based on Monk's work.}

The most comprehensive source of information on the development of the
CAD procedure and related work is \cite{caviness98}.%
\footnote{This includes a reprint of
Tarski's ``A decision method for elementary algebra and geometry'' \cite{tarski51}.
}
 Here are a few of the things I learned from that invaluable
volume.  In the first stage pursuant to quantifier elimination, the
CAD algorithm takes all the polynomials in the matrix of a prenex
formula $\phi$ with a total of $m$ free and bound variables, and
outputs a cell decomposition in $\mathbb{R}^m$, on each cell of which each of
the given polynomials is sign invariant; furthermore, the cells are
arranged in \emph{cylinders}.  The QE part of the algorithm uses the
output of the CAD algorithm to determine which cells of the
decomposition satisfy the matrix of $\phi$ in order to eliminate the
bound variables.  The first implementation of the CAD method was made
in 1979--80 by Collins' student Dennis S. Arnon.  In 1991, Collins and
another of his students, Hoon Hong, published a substantial
improvement for various examples in practice, though not in complexity
upper bound, requiring only \emph{partial} CAD  \cite{Collins_Hong:91}.  This was subsequently
implemented by Hong under the acronym QEPCAD.  The Caviness and Johnson
volume presents a number of applications, including polynomial
optimization, polynomial best approximation in lower degree (by
$n-2$ degree polynomials), the topology of semi-algebraic
sets, algebraic curve display, and robot motion planning.  By the way,
the system \emph{Mathematica} implements another form of CAD, according to
Adam Strzebonski of Wolfram Research, Inc. 

To round out the complexity picture, Fischer and Rabin \cite{fischer74} 
gave an EXPTIME lower bound of the
form $2^{cn}$ for deciding for sentences of length $n$ whether or not
they are true in the reals, no matter what algorithm is used; the same
applies even with non-deterministic algorithms, such as via proof
systems.  They also showed that the cut-point by which EXPTIME sets
in, i.e., the least $n_0$ such that for all inputs of length $n \ge
n_0$ , at least $2^{cn}$ steps are needed, is not larger than the
length of the given algorithm or axiom system for proofs.  Thus real
algebra is definitely infeasible on sufficiently large, though not
exceptionally large inputs.  The applications mentioned above are in a
gray area with relatively small numbers of variables, where
feasibility in practice depends on the specific nature of the problems
dealt with.  As for space complexity, there is a PSPACE lower bound on
the theory of the reals, as a consequence of a result of Stockmeyer's.    
Ben-Or, Kozen and Reif \cite{ben-or86} established an EXPSPACE upper bound and
conjectured that the set of true first-order sentences of the reals is
EXPSPACE-complete.  The exact time and space complexities of this set
is to this date an open problem (Phokion Kolaitis, personal
communication).\footnote{Just minutes before my lecture, I learned       
from Prakash Panangaden that John Canny (U.C. Berkeley School of
Engineering) proved \cite{Canny88b} that the existential theory of the reals
is in PSPACE.}

Tarski's own route to the decision problem for the reals began in the
mid-1920s with his development of an elegant first-order
axiomatization of geometry  \cite{tarski99}.
One of his main goals was to prove the completeness of this
axiomatization, and that led him to consider its interpretation in the
first-order theory of the reals.  Tarski recognized that the method of
eliminating quantifiers that had been initiated by Leopold L\"owenheim
and then applied by Thoralf Skolem and C. H. Langford was--when it
succeeded--a way of determining all the complete extensions of a
first-order axiom system--and in particular of proving
the completeness of complete systems.  In the latter part of the twenties      
Tarski ran the ``exercise sessions'' for the seminar at Warsaw
University led by the logic professor, Jan {\L}ukasiewicz, and he used
the opportunity to systematically pursue the method of elimination of
quantifiers.  As an ``exercise'', Tarski suggested to one of the
students, Mojzesz Presburger, that he find an
elimination-of-quantifiers procedure for the additive theory of
natural numbers, i.e., for the structure $\langle \mathbb{N}, +, <, 0,1\rangle$.
In that case, full quantifier-elimination is not possible, but can be
carried out in a definitional extension of its language , obtained by
adding as atomic formulas all those of the form $x \equiv y \pmod m$
 for each $m = 2, 3, 4, \ldots$. Mathematically, the procedure
comes down to solving a system of simultaneous congruences and thus
the Chinese remainder theorem.  Presburger's result served as his
master's thesis in 1928 and it was published a year later \cite{presburger29}. 
This slim paper of nine pages was to be his sole work in logic; after
that he went to work in the insurance industry.  Some people think
Presburger should have received the Ph.D. for that work, but it has to
be admitted that its significance was not realized until much later.%
\footnote{A sad coda to this story is
that Presburger, a Jew, perished in the Holocaust in 1943.}

The set of first-order truths of the additive structure of natural
numbers is called Presburger Arithmetic.  As I mentioned earlier,
Martin Davis presented his work on programming the Presburger
procedure on the {\em Johnniac}  at the Cornell conference in 1957.  That     
was long before Fischer and Rabin \cite{fischer74} showed that there is a
doubly-exponential time lower bound on any algorithmic procedure for
Presburger Arithmetic, including non-deterministic ones.  If Martin
had known that, he might not even have tried, even with today's
computers.\footnote{Shankar \cite{shankar02little} takes as an epigram a quote from
Davis \cite{Davis83} re his experiment with Presburger Arithmetic: ``Its great
triumph was to prove that the sum of two even numbers is even.''  A
second epigram from the same source, quoting Hao Wang, is that: ``The
most interesting lesson from these results is perhaps that even in a
fairly rich domain, the theorems actually proved are mostly ones which
call on a very small portion of the available resources\ldots''} On
the other hand, such lower bounds tell us little about the feasibility
in practice of deciding relatively short statements.  As to upper
bounds, Presburger's own procedure is non-elementary; this was
improved to triply-exponential by Derek Oppen \cite{oppen78}.  A search on
``Google Scholar'' came up with a number of references to Presburger
Arithmetic.  Near the top are applications to the symbolic model
checking of infinite state systems \cite{bultan97} and
proving safety properties of infinite state systems \cite{fribourg97};
 further applications via combination decision procedures
are indicated in \cite{shankar02little}.
	
Let's return to Tarski's own work on elimination of quantifiers for
the elementary (i.e., first-order) theory of real numbers: although it
was obtained by 1930 and he considered it to be one of his two most
important results (the other being his theory of truth), it's
surprising he didn't get around to preparing it for publication until
1939.  That was under the title, ``The completeness of elementary
algebra and geometry''%
\footnote{By the elementary theory of a
structure, Tarski means the set of its first-order truths.} 
for a new series on metamathematics planned by a Parisian publisher, but the
actual publication was disrupted by the German invasion of France in
1940. As Tarski later wrote: ``Two sets of page proofs which are in my
possession seem to be the only material remainders of that venture.''
The next time he got around to working on its publication was in 1948
when his friend and colleague J.~C.~C. McKinsey was at RAND
Corporation in Santa Monica.  My guess is that McKinsey suggested to
his superiors that there would be potential value to applying Tarski's
procedure to the computer calculation of optimal strategies in certain
games.  (Game theory was in those years a very popular subject at
RAND.)  However, any implementation would first require writing up its
theoretical details in full.  Working under Tarski's supervision,
McKinsey took on the job, revising the 1939 manuscript in its entirety.  That
came out as a RAND Report under the new title ``A decision procedure
for elementary algebra and geometry'' in 1948; it was finally brought
out publicly three years later by UC Press as a second edition \cite{tarski51}.  The
change in title from 1939 to 1948/1951 corresponds to a change in
aims, from completeness to decidability.  (By the way, a lightly
edited version of the 1939 page proofs eventually appeared under the
original title in 1967 in France).                                       

Though Tarski may not have been interested in actual computation at
any time in this entire history, he \emph{was} interested in
mathematical applications of his procedure.  In fact, one of Tarski's
strongest motivations throughout his career was to attract
mathematicians to the results of work in logic, and he often did this
by reformulating the results in a way that he thought would be more
digestible by mathematicians.  One side result he noticed about his
elimination-of-quantifiers argument for the first-order theory of the
real numbers is that every definable set has the form of a union of a
finite number of intervals (not necessarily proper) with algebraic
end-points.  He used this to illustrate the general concept of
definable set of elements in a structure.  At the outset of his 1931
paper on definable sets of real numbers \cite{tarski31} he said that mathematicians in
general don't like to deal with the notion of definability.  One
reason is that used informally it can lead to contradictions, like the
paradox of Richard; that uses an enumeration in English (say) of all
the real numbers definable in English, to define (in English) a real
number not in that enumeration, by diagonalization.  Another reason
for mathematicians' aversion mentioned by Tarski is that
mathematicians think the notion of definability is not really part of
mathematics.  In a way, he agrees, for he says that
\begin{quote}\small                             
The problems of making [the meaning of definability] more
 precise, of removing the confusions and misunderstandings connected
with it, and of establishing its fundamental properties belong to
\emph{another branch of science--metamathematics}.  [Italics mine]
\end{quote}
In fact, he says, he has ``found a general method which allows us to
construct a rigorous metamathematical definition of this notion''.

But then 
\begin{quote} \small                       
by analyzing the definition thus obtained it proves to be
possible\ldots to replace it by [one] formulated exclusively in
\emph{mathematical} terms.  Under this new definition the notion of
definability does not differ from other mathematical notions and need
not arouse either fears or doubts; it can be discussed entirely within
the domain of normal mathematical reasoning. [Italics mine]
\end{quote}
The \emph{metamathematical} explanation of definability in Tarski's
1931 paper is given in terms of the notion of satisfaction, whose
definition is only indicated there.  Under the \emph{mathematical}
definition, on the other hand, the definable sets and relations are
simply those generated from certain primitive sets of finite sequences
corresponding to the atomic formulas, by means of the Boolean
operations and the operation of projection.  Although the 1931 paper
concentrates on the concept of \emph{definability in a structure}, in a
footnote to the metamathematical explanation it is stated that ``an
analogous method can be successfully applied to define other concepts
in the field of metamathematics, e.g., that of \emph{true sentence\ldots}''

Tarski later spelled this out in his famous 1935 paper ``Der
Wahrheitsbegriff in den formalisierten Sprachen'' (The concept of truth
in formalized languages, \cite{tarski35}).\footnote{It was not until 1957, in a paper
with Robert L. Vaught \cite{tarski57}, that Tarski explicitly presented these notions
as those of satisfaction and truth in a structure.  See the discussion
by Hodges \cite{hodges85}  and Feferman \cite{feferman04b} of the relationship of that to
the ``Wahrheitsbegriff'' paper. } Some regard this work as one of the
most important instances of conceptual analysis in twentieth century
logic, while others think he was merely belaboring the obvious.  After
all, logicians like L\"owenheim, Skolem and G\"odel had been confidently
using the notions of satisfaction and truth in a structure in an
informal sense for years before Tarski's work and an explicit
definition was not deemed to be necessary, unlike, for example, the
conceptual analysis of computability by Turing.  I have to agree that
there is some justice to the criticism since the definitions are
practically forced on us, once one attends to providing them at all.
But even if that's granted, Tarski's explication of these concepts has
proved to be important as a paradigm for all the work in recent years
on the semantics of a great variety of formal languages.

In particular, the influence of Tarski on the semantics of programming
languages is so pervasive that to detail it would require an entire
presentation in itself.  Let me mention just one example, namely that
of the semantics of the lambda calculus and its extensions via domain
theory, as developed by Dana Scott and his followers.  This happens to
connect with the item in Tarski's list of publications that is most
cited in the computer science literature, namely his lattice-theoretic
fixpoint theorem \cite{tarski55}, which is an elegant abstract formulation of
the essential characteristic of definition by recursion.%
\footnote{Tarski proved that every monotonic function over a complete lattice has a complete lattice of fixed points, and hence a least fixed point.  This is a generalization of a much earlier joint result of Knaster and Tarski and so is sometimes referred to as the Knaster-Tarski theorem.  A related result used in applications is that every continuous function on a complete lattice has a least fixed point; credit for it is unclear, 
and thus it is               
considered a ``folk theorem''.  
The                          
history of these and other fixed point theorems relevant to computer science is surveyed in \cite{lassez82}.}  
There is also a significant personal connection: Scott began his studies in
logic at Berkeley in the early 50s while still an undergraduate.  His
unusual abilities were soon recognized and he quickly moved on to
graduate classes and seminars with Tarski and became part of the group
that surrounded him, including me and Richard Montague; so it was at
that time that we became friends.  Scott was clearly in line to do a
Ph.~D. with Tarski, but they had a falling out for reasons explained
in our biography of Tarski \cite{feferman04a}.  Upset by that, Scott left for Princeton where he    
finished with a Ph.D. under Alonzo Church.  But it was not long
before the relationship between them was mended to the point that
Tarski could say to him, ``I hope I can call you my student,'' and
rightly so: not only did Scott's thesis deal with a problem that had
been proposed by Tarski, but all of Scott's work is in the best
Tarskian tradition of breadth, rigor, clarity of exposition and
clarity of purpose.  And, like Tarski, he prefers set-theoretic and
algebraic methods, of which the domain-theoretic approach to the
semantics of type-free functional programming languages is a perfect
example.  So Tarski's influence on computer science manifests itself
here at just one remove, though of course Scott's contribution,
beginning in 1976 \cite{scott76} with the construction of a domain $D$ isomorphic to
$D\rightarrow D$, is completely novel.\footnote{Scott informed me that his use of lattice fixed points was initiated in the fall of 1969 in work with Christopher Strachey and exposed in many lectures in Oxford while on leave there.  For further developments and a large bibliography see \cite{scott03}.}

Satisfaction and truth in a structure are the basic notions of model
theory, whose systematic development in the 1950s is initially largely
due to Tarski and his school.  The notions are relative to a formal
language, which is usually taken to be first-order (FO), because of
the many happy properties of FO logic such as that of compactness
(cf. the texts by Chang and Keisler or Hodges).  But other kinds of                 
logics in which, e.g., compactness fails, turned out to be partially
susceptible to useful model-theoretic methods as shown by the greatly
varied contributions to the collection \emph{Model-Theoretic Logics} \cite{barwise85}.
 Among these are the model-theory of
infinitary languages as well as second-order and higher order
languages.  For computer science, a great variety of finite
structures, such as various classes of graphs, arise naturally, and it
was discovered that a number of questions in complexity theory may be
framed as questions in finite model theory (cf., e.g.                        
\cite{ebbinghaus91}).
In addition to first-order logic (FOL) and its finite
variable fragments, other logics that have proved to be useful in
finite model theory are finite-variable infinitary logics, monadic
second-order logic (MSOL) and its fragments, and certain fixed point
logics such as Datalog and least fixed-point logic LFP.  

Recently there has been a surge of very interesting work on analogues
in finite FO model theory to a class of general results called
preservation theorems in classical FO model theory.  The newest and
most exciting of these is due to Ben Rossman \cite{rossman05}.  So I can limit myself         
to explaining the general nature of the main results.  

Given a relation $R$ between structures and a sentence $\phi$, we say
that $\phi$ is preserved under $R$ if whenever $M$ satisfies
$\phi$ and $N$ is in the relation $R$ to $M$, then $N$ satisfies
$\phi$.  The results from classical FO model theory
characterize up to logical equivalence, the form of sentences
preserved under various $R$.  The most famous ones are the following,
all from the 1950s.

\begin{itemize}

\item EPT (\L os-Tarski).  $\phi$ is preserved under extensions iff
  $\phi$ is equivalent to an existential sentence.

\item OHPT (Lyndon).  $\phi$ is preserved under onto-homomorphisms iff
  $\phi$ is equivalent to a positive sentence.

\item HPT (\L os-Tarski-Lyndon).  $\phi$ is preserved under
(into-)homomorphisms iff $\phi$ is equivalent to an existential
positive sentence.
\end{itemize}

The finite analogues of these results are obtained by restricting to
finite $M$ and $N$.  The 'if' directions of course hold in all the
finite versions, but the 'only if' analogues of the Extension
Preservation Theorem (EPT) and the Onto-Homomorphism Preservation
Theorem (OHPT) are known to fail.  In particular, the failure of the
finite analogue of EPT is due to Bill Tait in 1959, who thereby
disproved a conjecture of Scott and Suppes; Tait's result was
rediscovered by Gurevich and Shelah in 1984.  The failure of OHPT in     
the finite is due to Rosen \cite{rosen02}.%
\footnote{Lyndon's famous positivity theorem implies OHPT.  Ajtai and Gurevich, and then     
Stolboushkin in a simpler way, proved failure in the finite of the positivity theorem, but their constructions did not prove failure in the finite of OHPT.} What Rossman \cite{rossman05} has proved,
surprisingly, is that HPT holds in the finite. There are interesting
relations to Datalog programs, which are given by existential positive
FO inductive definitions.  A Datalog formula may thus be considered as
an infinitary disjunction of existential positive FO formulas.  Using
a simple compactness argument, Rossman's result about HPT in the
finite also implies the theorem of Ajtai and Gurevich \cite{ajtai94} that on finite structures, if a Datalog sentence is equivalent to a FO sentence then it is equivalent to a single
existential positive sentence, exactly those preserved by
homomorphisms.%
\footnote{According to Rossman, the implication was
known to hold prior to his discovery.} 
The failure of the \L os-Tarski
characterization in the finite shows that preservation theorems do not
in general relativize from a class to a subclass.  Thus it is also of
interest to ask for which classes of finite structures HPT holds.
This had already been investigated by Atserias, Dawar and Kolaitis
\cite{atserias04} in which one of the main results is that HPT holds for every
class of finite structures of bounded treewidth; another result is
that HPT holds for the class of all planar graphs.

At the Tarski Centenary Conference held in Warsaw in 2001, Johann A. Makowsky presented a survey of applications of another kind
of preservation result that goes under the heading of the
Feferman-Vaught Theorem \cite{makowsky04}.  What Vaught and I had shown
in our joint 1959 paper \cite{feferman59} was that for a great variety of sum and
product operations O on structures $M_i\ (i \in I)$, the first-order
properties of 
$$M=O\langle M_i\ |\ i \in I \rangle$$ are determined by the first
order properties of each $M_i$ together with the monadic second-order
properties of a structure on the index set $I$.  It follows that
elementary equivalence between structures is preserved under such
operations $O$.  In later work, L\"auchli, Gurevich and Shelah extended       
our reduction-to-factors theorem to monadic second-order properties.
In his paper, Makowsky gives a unified presentation of this work with
emphasis on its algorithmic applications, in particular to splitting
theorems for graph polynomials.  I'll have to leave it at that, since
it would take too much time to try to go further into that here.

The final thing I want to tell something about is the connection of
Tarski's ideas and work with database theory.  Here it is not a matter
of direct influence but rather of the pervasiveness of his approach to
things, since the development of database theory apparently proceeded
quite independently.  Jan Van den Bussche has written an excellent
survey \cite{vandenbussche01applications} of the connections, which I urge you to read; here are a
few of the high points.  Codd \cite{Codd70} introduced a relational algebra
for expressing a class of generic (i.e., isomorphism invariant)
queries on databases; he also proved that the queries expressible in
his relational algebra are exactly those that are domain independent and definable in FO logic.  For
those who know Tarski's work on relation algebra, cylindric algebras
and algebraic logic more generally, the immediate question to raise is
the nature of the connection.%
\footnote{According to Van den Bussche
(personal communication) the first people from the database community
to recognize the connection between Codd's relational algebra and
Tarski's cylindric algebras were Witold Lipski and his student Tomasz
Imielinski, in a talk given at the very first edition of PODS (the ACM
Symposium on Principles of Database Systems), held in Los Angeles,
March 29-31, 1982.  Their work was later published in Imielinski and
Lipski \cite{imielinski84}.} 
(You can find a quick introduction to Tarski's work in
this respect in Interlude VI of our biography \cite{feferman04a}.%
\footnote{Some other     
applications to computer science--not discussed here--of Tarski's work
on relation algebra are indicated on p. 339 and its footnote 4 of that
interlude.}
) I view Tarski's work on algebraic logic as part of his
general effort to reformulate logic in mathematical as opposed to
metamathematical terms, in the hopes of thus making logic of greater
mathematical interest.  Tarski had done much work in the 1930s on
Boolean algebras, of which algebras of sets and algebras of
propositions (up to equivalence) are specific cases.  Stone's
representation theorem for Boolean algebras showed that every abstract
BA is isomorphic to a concrete one in the sense of fields of sets, and
in that sense the equational axioms of BA are complete.  With Tarski's
1941 paper ``On the calculus of relations'' \cite{tarski41} he single-handedly revived    
and advanced the $19^\mathrm{th}$ century work on binary relations by
Peirce and Schr\"oder, and introduced an elegant finite equational               
axiom system for relation algebra, from which all known special cases
of valid relational identities could be deduced.  However, it was
shown by Roger Lyndon in 1950 that there are non-representable                         
relation algebras, so Tarski's axioms are not complete; later, Donald
Monk proved (in 1964) that there is no finite axiomatization of the
valid equations in the language of these algebras.  This is related to
the fact that what can be expressed in relation algebra is exactly
what can be expressed about binary predicates in 3-variable
FOL.\footnote{Cf. the papers \cite{formisano04}, \cite{formisano05} dealing
with expressibility/inexpressibility in Tarski's algebraic
framework.}  See \cite{maddux91} for more on the history of relation algebras.           

Given the weakness of relation algebra, in the early 1950s Tarski
introduced the idea of \emph{cylindric algebras} (CAs) of dimension
$k$ for any $k \ge 2$, finite or infinite.  (\emph{NB}: Cylindric
Algebras have nothing to do with Cylindrical Algebraic Decomposition.)
In addition to the Boolean operations, these algebras use operations
$C_n$ of cylindrification for each $n < k$ and \emph{diagonal}
constants $d_{n,m}$ for each $n, m < k$.  The concrete interpretations
are given by fields of subsets of a $k$-ary space $U^k$, with the
$C_n$ interpreted as cylindrification along the $n$th axis, and the
$d_{n,m}$ as the set of $k$-tuples in $U$ for which the $n$th and
$m$th terms are equal.  Thus $k$-dimensional CAs abstract $k$-variable
FOL with identity.  The theory of CAs was extensively developed by
students and colleagues of Tarski and the results are exposited in the
volumes by Henkin, Monk and Tarski \cite{henkinvolume1},\cite{henkinvolume2}.  It turns out that
there are non-representable CAs for every $k \ge 2$, finite or
infinite, but Henkin and Tarski showed that ``locally finite'' CAs are
representable for every infinite $k$.  A CA is called locally finite if
for each element $a$ of the algebra, $C_n(a) = a$ for all but a finite
number of $n < k$. The local-finiteness condition
corresponds to each formula in FOL having at most a finite number of
free variables, and the representation theorem for infinite
dimensional locally finite CAs corresponds to the completeness theorem
for FOL with identity. 

As Van den Bussche points out in \cite{vandenbussche01applications},          
the language of $\omega$-CAs provides
an alternative to Codd's relational query language, and that of
$k$-CAs for $k$ finite is an alternative to queries definable in FOL with at most $k$ distinct variables.  But how
does Codd's language match up with that of Relation Algebras (RAs)?
In \cite{tarski87} it is shown that adjunction of a suitable           
``pairing axiom'' to RA makes it as strong as FOL. It turns out that the
corresponding idea has been developed in the case of database theory
by Gyssens, Saxton and Van Gucht \cite{Gyssens:94} using ``tagging'' operations,
giving a form ${\rm RA}^=$ that simulates Codd's relational
algebra.\footnote{Van den Bussche's article concludes with a survey
of some interesting connections to constraint databases and geometric
databases.}

\emph{So}, does that justify John Etchemendy's statement that the
shiny Oracle towers on Hwy 101 wouldn't be there without Tarski's
recursive definition of satisfaction and truth?  It would be more
accurate to say that the Oracle towers wouldn't be there without the
theoretical development of database theory, and \emph{that} wouldn't
be there without rethinking the model theory of first-order logic in
relation-algebraic and/or cylindric-algebraic terms, and \emph{that}
wouldn't be there without Tarski's promotion of both model theory and
algebraic logic.  Does Larry Ellison know who Tarski is or anything
about his work?  At the time of my lecture, I wondered whether Ellison
even knew who Codd was or the whole theoretical development of
database theory, without which the Oracle towers would indeed not be
there.  I learned subsequently from Jan Van den Bussche that not only
did Ellison know about Codd's work but he marks the reading of Codd's
seminal paper as the starting point leading to the Oracle Corporation;
cf. his biography given by the ``Academy of Achievement'' at               
\texttt{http://www.achievement.org/autodoc/page/ell0bio-1/}.
Actually, Codd himself didn't refer to Tarski in his fundamental
papers on database theory.  But other workers in the subject, such as
Imielinski and Lipski, and later, Kanellakis, did; they were well
aware of the connection and brought explicit attention to it (cf.,
e.g. \cite{kanellakis90}, p. 1085).  In whatever way the claim is
formulated, \emph{I} think it is fair to say that Tarski's ideas and
the approaches he promoted are so pervasive that even if his influence
in this and the various other areas of computer science about which I
spoke was not direct it was there at the base, and--to mix a                  
metaphor--it was there in the air, and so the nature and importance of
his influence eminently deserves to be recognized.

\bibliography{feferman}   

\end{document}